\documentstyle[preprint,aps]{revtex}             
\begin{document}
\preprint{ORNL-CTP-96-01,hep-ph/9604224} 
\draft

\title{ Effects of Screening on Quark-Antiquark Cross Sections in
Quark-Gluon Plasma}

\author{ Cheuk-Yin Wong$^1$ and Lali Chatterjee\cite{AAAuth}$^{1,2}$}

\address{ $^1$Oak Ridge National Laboratory, Oak Ridge, TN 37831}
\address {$^2$Department of Physics, University of Tennessee,
Knoxville, TN 37996} 

\date{\today}
\maketitle
 
\begin{abstract}

  Lowest-order cross sections for $q\bar q$ production and
annihilation can be approximately corrected for higher-order QCD
effects by using a corrective $K$-factor. For energies where quark
masses cannot be ignored, the $K$-factor is dominated by the wave
function distortion arising from the initial- or final-state
interaction between the quark and the antiquark.  We evaluate this
$K$-factor for $q \bar q$ production and annihilation in a quark-gluon
plasma by taking into account the effects of Debye screening through a
color-Yukawa potential.  We present the corrective $K$-factor as a
function of dimensionless parameters which may find applications in
other systems involving attractive or repulsive Yukawa interactions.
Prominent peaks of the $K$-factor occur for an attractive $q$-$\bar q$
color-Yukawa interaction with Debye screening lengths of 0.835 and
3.23 times the Bohr radius, corresponding to two lowest $s$-wave
$q\bar q$ bound states moving into the continuum to become $q\bar q$
resonances as the Debye screening length decreases.  These resonances,
especially the $c\bar c$ and the $b \bar b$ resonances, may be
utilized to search for the quark-gluon plasma by studying the
systematics of the temperature dependence of heavy-quark pair
production just above the threshold.

\end{abstract}


\narrowtext

\section{Introduction and Summary}

  The possible production of a deconfined quark-gluon plasma (QGP)
state during high-energy heavy-ion
collisions\cite{QM95,Won94,Shu88,Sat94,Mul95,Har96,Sta92} is
expected to be accompanied by Debye screening of the color charges of
the constituents in the plasma \cite{Mat86,Mat88,Kar88,Won94}. The
nature and the characteristics of the screening depend on the degree
of equilibration and the condition of the plasma.  Lattice QCD
calculations also provide information on the extent of color
screening\cite{Kar88}.  While the actual temperatures that might be
attained cannot be conclusively established at the present state of
the art, reasonable estimates of the temperature and other plasma
properties have been made for different scenarios and for different
experimental conditions\cite{Bjo83,Shu93,Kap94} . On the basis of
these, corresponding estimates can be made for the color screening
that would be operative for various scenarios.  Recent phenomenology
of soft-particle production \cite{Won96a,Won96b} and the ``hot glue''
partons scenario \cite{Shu93}, suggesting a possible initial
gluon-rich environment in nucleus-nucleus collisions, raise additional
questions regarding color screening in gluon-rich matter as well.

The screening phenomenon modifies the interaction between a quark and
an antiquark placed in the medium and affects their rates of
production and annihilation.  Many of these reactions, such as the
electromagnetic annihilation of $q$-$\bar q$ pairs and the production
of strange and charm quark pairs from the plasma, provide valuable
signals for the presence of the
plasma\cite{QM95,Won94,Shu88,Sat94,Mul95,Har96,Sta92}.  It is clearly
important to understand the influence of a screened color interaction
on various cross sections.

It is well known that the lowest-order $q\bar q$ annihilation and
production cross sections need to be corrected to take into account
initial-state and final-state interactions.  At high energies, a
$K$-factor is used to correct the tree-level rates to achieve a match
with experiments \cite{Kub80,Ham91,Fie89,Bro93,Gro86}.  At relatively
low energies near the $q$-$\bar q$ threshold, the importance of the
final-state interaction on the production of heavy quarks has been
recognized by many authors \cite{App75,Bar80,Gus88,Fad88,Fad90,Bro95}.
Similar effects of initial- and final-state interactions have also
been recognized in many other areas of
physics\cite{Gam28,Gyu81,Har57,Bet56}.  The Gamow factor (in nuclear
reactions \cite{Gam28}, pion production and pion interferometry
\cite{Gyu81}, and pair annihilations \cite{Har57}) and the $F$-factor
(in nuclear beta decay \cite{Bet56}) have been successfully utilized
to correct the lowest-order reaction cross sections or decay rates.
For example, in beta-decay, the probability for the occurrence of
low-energy electrons is greatly enhanced because of the attractive
final-state Coulomb interaction between the electron and the nuclear
charge, while the probability for the emission of low-energy positrons
is suppressed by the repulsive final-state Coulomb interaction.  The
corrective factors can be numerically quite different from unity,
especially at energies only slightly greater than the rest masses of
reacting participants.

   We have in an earlier work presented an approximate representation
for the $K$-factor to correct lowest-order $q\bar q$ cross sections to
take into account the initial- and final-state color-Coulomb
interaction without screening\cite{Cha95a}. At energies where quark
masses are important, we have determined the $K$-factor by the wave
function distortion effects resulting from the color-Coulomb
interaction between the quark and the antiquark. For the relativistic
case at very high energies, the corrective $K$-factor has been
determined from perturbative quantum chromodynamics (PQCD)
\cite{Kub80,Ham91,Fie89,Bro93}. Our approximate representation of the
corrective $K$-factor for the case of no screening interpolates
between the low- and the high-energy limits, following the
interpolation procedure of Schwinger \cite{Sch73}.  It has been tested
and found to give a good description of the E605 data on dilepton
production in the Drell-Yan process \cite{Mos95} and the experimental
ratio of $\sigma (e^+e^-\rightarrow$hadrons)/$
\sigma(e^+e^-\rightarrow\mu^+\mu^-)$ \cite{Cha95b}.

In this work we evaluate the $K$-factor by including the effects of
screening arising from the consideration of wave function distortion
only.  As these corrections are most important at low energies, we
restrict our investigation to non-relativistic equations of motion at
present. In future work, we hope to examine the relativistic case
where the relative energy is much greater than the quark masses.

Attractive and repulsive Yukawa potentials appear in many other areas
of physics.  To make the corrective $K$-factor obtained here useful
also for reactants in any attractive or repulsive Yukawa 
potential, we present the corrective factor as a function of two
dimensionless parameters.  For attractive screened interactions, we
observe interesting peaks of the $K$-factor as a function of the
screening length, which correspond to single-particle $q \bar q$ bound
states moving into the continuum as the screening length decreases.
These peaks occur when the Debye screening length $\lambda_{{}_D}$ is
equal to 0.835 and 3.23 times the Bohr radius $a_{{}_B}$.

The phase shifts and scattering cross sections for the attractive
screened Coulomb potential have been evaluated earlier by Morse and
his collaborator for the scattering of electrons from an
atom\cite{Mor32,All31} and also by Calogero\cite{Cal67}.  By comparing
our $K$-factor results with the phase-shifts of Morse and
collaborators, we found that the $s$-wave resonances at $E\approx 0$
occur at the same ratios of $\lambda_{{}_D}/a_{{}_B}$ as the peaks of
the $K$-factor.  Therefore, the occurrence of the peaks of the
$K$-factor and the resonances at $E\approx0$ are the manifestations of
the same physical phenomenon arising from emergence of the
single-particle $q\bar q$ bound states into the continuum to become
$q\bar q$ resonances.  Because they arise from Debye screening in a
deconfined quark-gluon plasma at high temperatures, they can be called
``screening resonances'' to be distinguished from other usual $q\bar
q$ resonances such as $J/\psi$, $\psi'$, and $\Upsilon$ in the
confining phase.

The occurrence of the $q\bar q$ screening resonances, especially the
$c\bar c$ and the $b \bar b$ screening resonances, may provide a way
to search for the quark-gluon plasma.  Because the screening length is
inversely proportional to the temperature, the peaks in the $K$-factor
will appear at specific values of the temperature.  Therefore, very
large enhancement of the production of $c\bar c$ or $b\bar b$ pairs
near the thresholds will occur at the corresponding temperatures
specified by the resonance screening length parameter.  A systematic
study of the dependence of the $K$-factor for heavy-quark pair
production near the threshold as a function of temperature will be
useful to map out the effect of screening, which will identify
important characteristics of the deconfined quark-gluon plasma.

This paper is organized as follows.  In Section II, we discuss the
phase-amplitude method to calculate the $K$-factor for a screened
Coulomb-type potential.  The corrective $K$-factor is calculated for
attractive and repulsive potentials and the results are exhibited in
Section III as a function of two dimensionless parameters.
In Section IV, we calculate the $K$-factors for charm and strange pair
production in a quark-gluon plasma. We show how the screening modifies
the lowest-order cross sections for $s\bar s$ and $c\bar c$ production
in Section V.  Section VI summarizes the present investigation and
discusses how the $q\bar q$ resonances in the deconfined
phase may be used to search for the quark-gluon plasma.

\section{The $K$-factor for Screened Color Potential }

Basic processes involving $q$-$\bar q$ initial-state or final-state
interactions include dilepton production through $q\bar q \rightarrow
\gamma^* \rightarrow l\bar l$, charm and strange pair production
through $q \bar q \rightarrow g^* \rightarrow q \bar q~ (s \bar s,
c\bar c)$ and $g g \rightarrow q \bar q ~ (s \bar s, c\bar c)$.

In the lowest-order reaction cross sections for these processes, the
quark and the antiquark are described by plane waves, and their mutual
initial- or final-state interactions are not included.  The quark and
the antiquark however interact with each other through a color-Coulomb
strong interaction, with a large coupling constant.  While one can
obtain the effect of these interactions to the next order in
perturbation theory, there are situations in which a perturbation
expansion involving only the lowest two orders may not be sufficient,
as in the case of large coupling constants and at energies where the
quark masses are not negligible.  A nonperturbative correction, based
on the use of the distorted wave function in the presence of the
color-Coulomb potential, can provide non-perturbative corrections to
the cross sections \cite{Gus88,Fad88,Fad90,Bro95,Cha95a}.

If one places a quark and an antiquark in a color medium such as a
quark-gluon plasma or gluon-rich matter, the interaction between the
quark and the antiquark will be screened.  In the case of a
quark-gluon plasma, the degree of screening depends on the
temperature.  The greater the temperature, the greater the density of
quarks, antiquarks, and gluons, and the greater will be the degree of
screening.  The effects of Debye screening in the quark-gluon-plasma
will modify the configuration space representation of the virtual
gluon propagator between the quark and the antiquark.  The color
potential connecting the quark and the antiquark will be modified from
a color-Coulomb interaction into a color-Yukawa interaction. (For an
introduction, see page 349 of Ref. \cite{Won94}.)  The screened 
interaction will modify the quark-antiquark relative wave function,
and will therefore change the corrective $K$-factor which is given by
the square of the wave-function at contact.  We therefore need to
evaluate the wave function for a quark and an antiquark in a screened
color potential.

The $q$-$\bar q$ Schr\"odinger equation in the relative coordinate $r$
can be written as
\begin{eqnarray}
\label{eq:Sch}
[\nabla ^2 + k^2 - 2 \mu  V(r)]\psi (r) =0\,,
\end{eqnarray}
where $\mu$ is the reduced mass, and the wave number $k$ at infinite
separation is related to the nonrelativistic kinetic energy $E$ by
$k^2=2\mu E$.  The screened color potential $V(r)$ which describes the
interaction between the quark and the antiquark can be written
as\cite{Won94,Mat86,Mat88}
\begin{eqnarray}
\label{eq:pot}
V(r)=-{ \alpha_{\rm eff}  e^{-r/\lambda_{{}_D}} \over r} \,,
\end{eqnarray}
where the Debye screening length $\lambda_{{}_D}$ is inversely
proportional to quark-gluon plasma temperature $T$ \cite{Gro81}, and
the effective coupling constant $\alpha_{\rm eff}$ is related to the
strong coupling $\alpha_{s}$ by the color factor $C_f$:
\begin{eqnarray}
  \alpha_{\rm eff}={C_f\alpha_{s} } \,.
\end{eqnarray}
We have included a negative sign on the righthand side of the
potential (\ref{eq:pot}). Thus, when $q$ and $\bar q$ are in a 
color-singlet state, $C_f$ is $4/3$ \cite{Fie89}, $\alpha_{\rm eff}$
is positive, and the interaction is attractive.  When $q$ and $\bar q$
are in a color-octet state, $C_f$ is $-1/6$ \cite{Fie89},
$\alpha_{\rm eff}$ is negative, and the interaction is repulsive.

Following Calogero\cite{Cal67}, we use the phase-amplitude
method to solve the Schr\"odinger equation for the $s$-state.  We
write the $s$-wave radial wave function $u_0(r)=r \psi(r)$ in terms of the
phase shift function $\delta_0(r)$ and the amplitude function
$\alpha_0(r)$ as
\begin{eqnarray}
u_0(r)=\alpha_0(r) \biggl \{ \cos\delta_0(r) ~[kr j_0(kr)] - \sin \delta_0(r)~
[kr \eta_0 (kr)] \biggr \}
=\alpha_0(r) \sin(kr +\delta_0(r) ).
\end{eqnarray}
From the Schr\"odinger equation (\ref{eq:Sch}), the equation for the
$s$-wave phase shift function $\delta_0(r)$ is \cite{Cal67}
\begin{eqnarray}
\label{eq:difeq}
{d \delta_0(r) \over dr}= - k^{-1} 2 \mu V(r)  \sin^2 [kr + \delta_0(r)].
\end{eqnarray}
This is a first-order differential equation which can be
solved numerically starting from the origin, with the boundary condition
$\delta_0(0)=0$.  After the function $\delta_0 (r)$ is determined, the
amplitude function $\alpha_0(r)$ can be obtained from 
the relation \cite{Cal67}
\begin{eqnarray}
\label{eq:in}
\alpha_0(r) = \alpha_0(0)~ \exp \biggl \{ (2 k) ^{-1} \int_0^r dr' [2
\mu V(r')] \sin 2 [k r' +\delta_0 (r')] \biggr \}\,.
\end{eqnarray}
We then determine the corrective factor $K$ from the amplitude
function at large $r$ using
\begin{eqnarray}
K = \biggl | {\alpha_0(r=0) \over \alpha_0(r \rightarrow \infty)
}\biggr | ^2 \,.
\end{eqnarray}
In terms of the phase shift function $\delta_0(r)$, the corrective
$K$-factor is given by 
\begin{eqnarray}
K =  \exp \biggl \{ -{1 \over k}  \int_0^\infty dr' [2
\mu V(r')] \sin 2 [k r' +\delta_0 (r')] \biggr \} \,.
\end{eqnarray}
It is convenient to express all radial distances in units of the Bohr
radius
\begin{eqnarray}
a_{{}_B}={1\over \mu |\alpha_{\rm eff}| }\,.
\end{eqnarray}
The Schr\"odinger equation in terms of $x=r/a_{{}_B}$ contains only
the usual `Coulomb' parameter
\begin{eqnarray}
\xi= {\alpha_{\rm eff} \over v} 
\end{eqnarray}
where $v=k/\mu$ and $\xi$ can be positive or negative, and the
dimensionless screening length parameter
\begin{eqnarray}
\eta={ \lambda_{{}_D} \over a_{{}_B}} \,.
\end{eqnarray}
The $K$-factor is therefore only a function of $\xi$ and $\eta$; we
shall show results for 
$K(\xi,\eta)$ in the next section.  

Note that the nonrelativistic energy $E$ is equal to $1/(2\mu
a_{{}_B}^2 \xi^2)$, and the wave number $k$ is equal to $1/(a_{{}_B}
|\xi|)$; thus, $E$ and $k$ decrease as $|\xi|$ increases.  In
studying $K(\xi,\eta)$ for the quark-gluon plasma, one can consider 
$|\xi|$ a measure of the inverse of $\sqrt{E}$ and $\eta$ a measure
of the inverse of the quark-gluon plasma temperature $T$.

\section{ Features of the Corrective Factor for Screened Potentials}

To illustrate some general features of the corrective factor
$K(\xi,\eta)$, it is instructive to consider a simple potential
similar to the screened potential (\ref{eq:pot}), for which analytical
corrective factor can be easily obtained.  The truncated color-Coulomb
potential ${\tilde V}(r)=-\theta(a-r) \alpha_{\rm eff}/r$ is
qualitatively similar to the screened Yukawa potential (\ref{eq:pot}),
and can be chosen to match the Yukawa potential (\ref{eq:pot}) with
the same volume integral by setting $a=\sqrt{2}\lambda_{{}_D}$.
It can be shown easily that the corrective $K$-factor for this
truncated color-Coulomb potential is
\begin{eqnarray}
\label{eq:kval}
  K= \biggl | {\sin (ka+\delta_0(a))  \over F_0(\xi, ka)} \biggr |^2 
{ 2 \pi \xi  \over 1- \exp \{ {-2 \pi \xi } \} } ,
\end{eqnarray}
where $\delta_0(a)$ is the phase shift for the truncated color-Coulomb
potential ${\tilde V}(r)$ and $F_0(\xi,ka)$ is the usual regular
$s$-wave Coulomb wave function (Eq.\ (14.1.3) of Ref.\ \cite{Abr65}).

For large values of $a$ and $ka>>1$, we have $\delta_0(a)=\xi\ln(2ka)+ \arg
\Gamma(1-i\xi)$ and 
\begin{eqnarray}
\label{eq:kval1}
K= \biggl | {\sin \theta_0 \over f \sin \theta_0 + g \cos \theta_0
}\biggr |^2 { 2 \pi \xi \over 1- \exp \{ {-2 \pi \xi } \} } ,
\end{eqnarray}
where 
\begin{eqnarray}
\theta_0=ka+\xi \ln(2ka) +\arg \Gamma(1-i\xi)\,,
\end{eqnarray}
and the asymptotic expansion of $F_0(\xi,ka)$ gives\cite{Abr65} 
\begin{eqnarray}
f=1-{\xi\over 2 ka} + {5\xi^2 - \xi^4 \over 8 k^2 a^2}\,,
\end{eqnarray}
\begin{eqnarray}
g={\xi^2 \over 2 ka} - {4\xi^3 - 2\xi \over 8 k^2 a^2}\,. 
\end{eqnarray}
The first factor on the righthand side of Eq.\ (\ref{eq:kval1}) is
close to unity, with small oscillatory corrections.  Therefore, for
large values of $a$ (which is proportional to $\eta$), $K(\xi,\eta)$
approaches and oscillates about the Gamow factor
$2\pi\xi/(1-\exp\{-2\pi\xi\})$, provided that $ka$ is also large,
corresponding to $\eta>>|\xi|$.

For small values of $a$ (or $\eta$) and $ka$, we can evaluate the phase shift
$\delta_0(a)$ in the Born approximation and obtain
\begin{eqnarray}
\delta_0(a)=\xi k^2 a^2(1- {k^2 a^2 / 6}) \,.
\end{eqnarray}
From Eq.\ (\ref{eq:kval}), an expansion of the Coulomb wave function in powers of $ka$ leads to
\begin{eqnarray}
\label{eq:kval2}
K &=&{[1 +\xi ka( 1- k^2 a^2/6)- k^2 a^2/6  ]^2
\over [1 - \xi k a -(2 \xi^2-1)k^2 a^2/6 ]^2  } \nonumber\\
 &\approx & { 1 + 2 \sqrt{2}~ {\rm sign}(\xi) \eta \over 1 - 2 \sqrt{2}
~{\rm sign}({\xi}) \eta } \,,
\end{eqnarray}
where ${\rm sign}(\xi)$ is the sign of $\xi$.  By our convention in Eq.\
(\ref{eq:pot}), ${\rm sign}(\xi)=1$ for attractive $q$-$\bar q$
interactions and ${\rm sign}(\xi)=-1$ for repulsive $q$-$\bar q$ interactions.
Therefore, for small values of $\eta$ and small $ka$  (which
corresponds to $|\xi| >>\eta$), $K(\xi,\eta)$ is approximately unity
and is independent of $|\xi|$. 

Having discussed the limiting behavior of $K(\xi,\eta)$, we return to
the screened color-Coulomb potential of Eq.\ ({\ref{eq:pot}) and
display the corresponding corrective factor $K(\xi,\eta)$ as a
function of $\xi$ and $\eta$.  The corrective factor $K(\xi,\eta)$ for
$-1 \le \xi \le 1$ and $\eta=0.1, 0.2,...,0.7$ is shown in Fig.\ 1.
It is greater than unity for positive $\xi$, corresponding to an
attractive interaction pulling the interacting particles together.  It
is less than unity for negative $\xi$, corresponding to a repulsive
interaction pushing the interacting particles apart.  For small
$\eta$, it approaches unity, indicating the diminishing influence of
the interaction between the reacting particles for small screening
lengths. The values of $K(\xi,\eta)$ for small $\eta$ and large $\xi$
agree with the estimates from Eq.\ (\ref{eq:kval2}).

We can also compare our results to the limit of no screening.  It may
be recalled that in this limit with $\eta \rightarrow \infty$, the
corrective factor $K(\xi,\eta)$ approaches the Gamow factor which is
only a function of $\xi$.  The Gamow factor is shown in Fig.\ 1 as the
dashed curve.  For the range $0\le \xi\le 1$ shown in Fig.\ 1,
$K(\xi,\eta)$ increases as $\xi$ and $\eta$ increase.  It is quite
close to the Gamow factor for $\eta > 0.7$.

In Fig.\ 2 we show $K(\xi,\eta)$ as a function of $\xi$ in the range
$1 \le \xi \le 3$ for various values of the screening length parameter
$\eta$.  The Gamow factor is the dashed curve.  For small $\eta$,
$K(\xi,\eta)$ is nearly constant in $\xi$.  As the screening length
parameter $\eta$ increases, the corrective $K$-factor increases to
become greater than the Gamow factor, reaches a maximum value, then
reverses itself and oscillates about the Gamow factor.  This
oscillation about the Gamow factor can be understood from the
approximate estimate of Eq.\ (\ref{eq:kval1}).

The peculiar behavior of the rise of $K(\xi,\eta)$ in the region of
$\eta\approx 1$ merits closer scrutiny.  Accordingly, in Fig.\ 3a we
study $K(\xi,\eta)$ as a function of $\eta$ in the range $0.4 \le \eta
\le 2$.  We find that $K(\xi,\eta)$ is a maximum at $\eta=0.835$.  The
larger the values of $\xi$, the greater is the maximum value of
$K(\xi,\eta)$.  The peak values of the $K$-factor are much greater
than the corresponding Gamow factor, as indicated by the ratio
$K(\xi,\eta)/$(Gamow factor) as a function of $\eta$ in Fig.\ 3b.

The prominent peak of $K(\xi,\eta)$ at $\eta=0.835$ is due to the
emergence of the lowest bound state of the system into the continuum
to become a $q\bar q$ resonance as the screening length
decreases. Movement of bound states into the continuum is familiar in
the context of a single-particle system in a finite-range potential,
as in the scattering of an electron from an atom\cite{Mor32,All31}. An
attractive screened potential with a large screening length is able to
hold many negative-energy bound states.  As an illustration, consider
an attractive Coulomb potential which holds an infinite number of
bound states.  As the screening length $\lambda_{{}_D}$ decreases from
infinity, the Coulomb potential becomes a Yukawa potential with a
decreasing range and its bound states are pushed into the continuum.
Using semiclassical estimates, it has been noted that when the
screening length $\lambda$ is comparable to the Bohr radius, the
screened potential cannot hold any bound state \cite{Mat88,Won94}.
Numerical calculations show that $\eta\approx 0.84$ is the screening
length parameter for which the lowest bound state in the screened
potential becomes unbound \cite{Mat88}.  This coincides with the
location of the peaks of $K(\xi,\eta)$ in $\eta$.  Combining the
bound-state result of Ref.\ \cite{Mat88} with our continuum result for
the $K$-factor, we conclude that the peak of the corrective factor at
$\eta=0.835$ arises because the lowest $q\bar q$ bound state emerges
into the continuum to become a $q\bar q$ resonance as the screening
length decreases.  The occurrence of similar resonances for electron
scattering from an atom has been observed earlier by Morse and
collaborators\cite{Mor32,All31}.

There are other well-known physical phenomena in which bound states
emerge into the continuum to become resonances as the range of the
attractive potential decreases.  For example, in nuclear physics the
$s$-wave strength function obtained from very low energy neutron
scattering shows peak values as a function of the nuclear mass number
at $A \approx 60$ and $A\approx 160$. (See page 230 of Ref.\
\cite{Boh67}.) The $s$-wave strength function is related to the wave
function at the nuclear surface, which is correlated with the wave
function at the nuclear interior.  The peaks of the $s$-wave strength
function at $A \approx 60$ and $160$ correspond to the emergence of
the bound single-particle $4s_{1/2}$ and $3s_{1/2}$ states
respectively into the continuum as the nuclear mass number decreases.
(See page 239 of Ref.\ \cite{Boh67}.)  The decrease of the nuclear
mass number is accompanied by a decrease of the range of the
attractive interaction.  As a consequence, nucleon single-particle
states are pushed into the continuum as the range of the interaction
decreases, similar to the emergence of $q\bar q$ bound state into the
continuum as the screening length decreases, as found in the present
study.

The behavior of the $q\bar q$ resonance at $\eta=0.835$ as a function
of energy can be studied by examining the asymptotic phase shift
($\delta_0\equiv \delta_0(r\rightarrow \infty)$) for the screened
potential as a function of $E$ (in units of $(2\mu a_{{}_B}^2)^{-1}$).
For $\eta>0.835$, there is a bound state and the asymptotic phase
shift $\delta_0$ starts at $\delta_0=\pi$ at $E=0$ (according to
Levinson's Theorem \cite{Lev49}), and $\delta_0$ decreases
monotonically as the energy increases (as in the short-dashed curve in
Fig.\ 4).  At $\eta=0.835$, the asymptotic phase shift starts at
$\delta_0=0$ at $E=0$, rises to approximately $\pi/2$, and then
decreases as the energy increases (the solid curve in Fig.\ 4).  When
$\eta < 0.835$, $\delta_0$ rises from zero to a value close to $\pi/2$
and decreases at higher energies.  As $\eta$ decreases, the resonance
peak (the location of the maximum $\delta_0$) moves to higher
energies, the maximum phase shift decreases, and the width of the
structure increases.  For $\eta < 0.835$, although the phase
shift nearly passes through $\pi/2$ the corrective $K$-factor does
not show a prominent peak structure at the resonance energy (the
location of maximum $\delta_0$), because the corrective $K$-factor is a
rapidly decreasing function of the energy (approximately as $1/\sqrt{E}$).
The best indicator for the presence of this $q\bar q$ resonance at
$\eta=0.835$ appears to be the the prominent peak structure of the
corrective $K$-factor as a function of $\eta$, much as the
manifestation of $s$-wave resonances from the $s$-wave strength
function plotted as a function of the nuclear radius or the nuclear
mass number.

It is interesting to inquire whether there are similar peaks of
$K(\xi,\eta)$ at other locations of $\eta$.  We have found that there
is another similar peak of $K(\xi,\eta)$ at $\eta=3.23$ (Fig.\ 5)
corresponding to the second $s$-wave bound state emerging into the
continuum to become a $q\bar q$ resonance.  It is clear that other
peaks of $K$-factor can occur at higher values of $\eta$ when
single-particle $s$-wave bound states are pushed into the continuum to
become $s$-wave $q\bar q$ resonances as $\lambda_{{}_D}$ decreases.

The $K$-factor for negative values of $\xi$ gives the corrective
factor for the lowest-order cross section involving two particles
subject to a mutually repulsive screened interaction, as in the case
of a $q$ and a $\bar q$ in color-octet states or two $\pi^+$ particles
subject to a screened Coulomb repulsion in pion interferometry.  In
Fig.\ 6 we show $K(\xi,\eta)$ as a function of $\xi$ in the range $-3
\le \xi \le 0$ for different screening length parameters
$\lambda_{{}_D}$=$\{0.1,0.6,..,3.1\}$.  The corresponding Gamow factor
(the no-screening limit) is also displayed.  As one observes in this
range of $\eta$, the $K$-factor for the screened potential is greater
than the Gamow factor, and the deviation from the Gamow factor becomes
greater, the larger is $|\xi|$.  The screening length $\eta$ need to
be much greater than $|\xi|$ to reach the no-screening limit.  Thus,
the effect of screening is also important for repulsive screened
potentials, and the use of the Gamow factor for the case with
screening may sometimes greatly over-correct the initial- or
final-state interactions if $\eta$ is small.

\section{ Correction Factors  for Typical Plasma}

In order to study the case of screening in the quark-gluon plasma, we
make a minor modification to the general case examined in the last two
sections.  While we use the non-relativistic potential (\ref{eq:pot})
to describe the interaction between a quark and an antiquark, we would
like to include relativistic kinematics for the system, as we consider
also kinetic energies with associated velocities of the order of
$0.5$.  The use of relativistic kinematics will avoid errors due to
the neglect of the higher powers of the momentum, which would be
present if we restricted ourselves to non-relativistic energy-momentum
relations.  In place of Eq.\ ({\ref{eq:Sch}), the equation of motion
to be solved is \cite{Cha95a,Cra92}
\begin{eqnarray}
\label{eq:69}
\biggl( - {\bf p}^2 + b^2 -2 \epsilon_\omega A_0  \biggr ) \psi  = 0
\end{eqnarray}
where $A_0(r)=V(r)$, and the quantities $b$ and
$\epsilon_\omega$ can be expressed in terms of the center-of-mass
energy $\sqrt{s}$ as
\begin{eqnarray}
\epsilon_\omega=(s-m_1^2 -m_2^2)/2 \sqrt{s}
\end{eqnarray}
\begin{eqnarray}
b^2=[s^2 -2 s (m_1^2 +m_2^2)+(m_1^2-m_2^2)^2]/4 s
=\epsilon_\omega^2-m_w^2
\end{eqnarray}
where $m_w=m_1 m_2 /\sqrt{s}$.  
The corresponding dimensionless parameters $\xi$ and $\eta$ are
\begin{eqnarray}
\xi={\alpha_{\rm eff} \over v}
={\alpha_{\rm eff} \epsilon_\omega \over b}
\end{eqnarray}
and
\begin{eqnarray}
\eta={\lambda_{{}_D} \over a_{{}_B}}
={\lambda_{{}_D} \alpha_{\rm eff} \epsilon_\omega }\,.
\end{eqnarray}
Values of $1.5$ GeV and $0.15$ GeV have been used for the masses of
the charm and strange quarks respectively.

The method outlined in Section III allows us to calculate the
distortion corrections, which comprise the most important contribution
to the $K$-factor at energies where the quark masses are not
negligible.  At this time we do not interpolate the $K$-factor to
high-energy results involving radiative corrections; the finite
temperature radiative corrections derived for the quark-gluon
plasma\cite{Bra90,Alt89} may be useful for this purpose.

For the quark-gluon plasma with a flavor number $N_f$ and $N_c=3$,
lowest-order perturbative QCD gives a Debye screening length of
\cite{Gro81}
\begin{eqnarray}
\lambda_{{}_D}({\rm PQCD})
=
{1 \over \sqrt{  \bigl ( {  N_c \over 3}+ {N_f \over 6 }\bigr ) g^2  }
~T} \,.
\end{eqnarray}
For a coupling constant $\alpha_s=g^2/4\pi=0.3$ and $N_f=3$, the Debye
screening length at a temperature of 200 MeV is $\lambda_{{}_D}
\approx 0.4$ fm.  We examine the cases of $\lambda_{{}_D}=0.2$ and 0.4
fm, corresponding respectively to temperatures of 400 MeV and 200 MeV
in this perturbative QCD estimate.  We will calculate the associated
$K$-factors for charm and strange pairs.

 We show in Fig.\ 7a the corrective $K$-factor for $c\bar c$ in a 
color-singlet state for which $C_f=4/3$ and $K(\xi,\eta) >1$.  
For the $c$-$\bar c$ interaction, we use the running coupling constant
\cite{Fie89,Bro93}
\begin{eqnarray}
\label{eqn:alp}
\alpha_{s}(Q^2) = { 12 \pi \over (33 - 2 N_f ) \ln (Q^2/\Lambda^2) },
\end{eqnarray}
and we assume four flavors and $\Lambda_{QCD} = 300$ MeV.  For
$\lambda_{{}_D}=0.4$ fm, the $K$-factor is much greater than unity
near the threshold of 3 GeV, and it decreases rapidly as a function of
energy.  The effective coupling constant is approximately $\alpha_{\rm
eff}= 0.4$ in this energy region, which gives a Bohr radius of
$a_{{}_B} \approx 0.5$ fm.  Thus, the screening length of
$\lambda_{{}_D}=0.4$ fm corresponds to $\eta \approx 0.8$, which is
near $\eta=0.835$ of peak $K$-factors, as we can infer from Fig.\ 3.
The $K$-factor oscillates about the Gamow factor (the no-screening
limit) which is shown as the dashed curve.  It is much greater than
unity near the threshold.  For $\lambda_{{}_D}=$ 0.2 fm the $K$-factor
is not as large.  This corresponds to a value of $\eta\approx$ 0.4 for
this energy region, far from $\eta=$ 0.835 where the peak of the
$K$-factor is located.

We show in Fig.\ 7b the corrective $K$-factor for $c\bar c$ in the
color-octet state, for which $C_f=-1/6$ and $K(\xi,\eta)<1$.  It
decreases with increasing $\lambda_{{}_D}$ and varies relatively
smoothly with energy.  The Gamow factor is also shown for comparison.
The color-octet corrective factor is greater than the Gamow factor, and
its deviation from unity is less than the deviation of the Gamow
factor from unity.

Because the magnitude of the octet repulsive potential is less than
the magnitude of the singlet attractive potential, the deviation of
the $K$-factor from unity is greater for color-singlets than for
color-octets.

We have extended our study to strange pair production or annihilation,
as strange pair production is an important process in high-energy
heavy-ion collisions which can be used as a signature of a QGP.  To
study strangeness production in the quark-gluon plasma, we use the 
running coupling constant proposed by Godfrey and Isgur\cite{Isg85},
which is appropriate for the energy range near the strange quark pair
threshold:
\begin{eqnarray}
\alpha_s(Q^2) = 0.25 \exp(-Q^2) + 0.15 \exp(-Q^2/10) +0.2 \exp(-Q^2/1000),
\end{eqnarray}
where $Q$ is in GeV.  We display in Figures 8($a$) and 8($b$) the $s
\bar s$ corrective $K$-factor in the presence of screening for
color-singlets and color-octets respectively.

 An interesting difference is observed in the color-singlet $K$-factors
for charm and strange pairs near threshold for the same values for the
screening length. For a charm quark pair, the corrective $K$-factor is
larger than for a strange pair, and increases with decreasing
energy. On the other hand, the color-singlet $K$-factor for strange
quark pairs decreases slowly with decreasing energies. Both systems
are studied at similar velocities as we are considering regions
in which  quark masses are important. Although the running coupling
constants are different because  the two systems sample different energy
regions at threshold, this is not sufficient to account for the
difference.

The difference in $K(\xi,\eta)$ for the charm and strange flavors in a
similar velocity ranges can be understood in the light of the analysis
in Section III. For screening lengths of $0.2$ fm and $0.4$ fm
considered here, the values of $\eta$ in the $c\bar c$ system are
$\eta\approx$ 0.4 and 0.8 respectively.  The corresponding
$K(\xi,\eta)$ have been discussed.  For the $s\bar s$ system on the
other hand, these values of $\lambda_{{}_D}$ correspond to much
smaller values of $\eta$, due to the smaller mass of the strange
quark. We find $\eta\approx$ 0.04 and 0.08 for the two $s\bar s$
cases, and the deviation of $K(\xi,\eta)$ from unity is small. This is
in accordance with the physical expectation that the interaction is
suppressed when the range of the potential is small compared to the
natural length scale of the system $a_{{}_B}$, so the $s \bar s$
system responds almost as an unperturbed one.

\section {Strange and charm quark-antiquark production}

We now consider the effects of screening on $c\bar c$ and $s \bar s$
production cross sections near and above the threshold.  The basic
cross section for $q\bar q \rightarrow g^* \rightarrow Q\bar Q$,
(where $Q$ refers to $s$ or $c$ quarks), averaged over initial and
summed over final colors and spins can be written as\cite{Com79}
\begin{eqnarray}
\sigma_{q\bar q} (M_{Q\bar Q}) =   { 8 \pi\alpha^2_s \over 27 M_{Q\bar
Q}^2}
\biggl (1+{\eta_m
\over 2}\biggr) \sqrt{1-\eta_m}\,,
\end{eqnarray}
where $\eta_m =4 m^2_Q/M_{Q\bar Q}^2$, $m_Q$ is the mass of
the quark and $M_{Q\bar Q}$ is the invariant mass of the
produced $Q\bar Q$ pair.

The corresponding expression for the gluon fusion mode, averaged
over initial gluon types and polarizations and summed over final
colors and spins is \cite{Com79,Won94}
\begin{eqnarray}
\sigma_{gg} (M_{Q\bar Q}) = { \pi\alpha^2_s \over 3 M_{Q\bar
Q}^2}\biggl\{ (1 +\eta_m +{1\over 16} \eta_m^2) \ln
\biggl({1+\sqrt{1-\eta_m}\over 1-\sqrt{1-\eta_m}} \biggr) - \biggl(
{7\over 4} + {31\over16} \eta_m \biggr) \sqrt{1-\eta_m}
\biggr)\biggr\}\,.
\end{eqnarray}
To obtain the overall $Q \bar Q$ ($s\bar s$ and $c\bar c$) production,
the above cross sections must be convoluted with the appropriate quark
and gluon distributions.  For charm and strange pair production in the
quark-gluon plasma, these tree-level cross sections must be corrected
to take into account the effect of Debye screening.

For $Q\bar Q$ production by quark-antiquark annihilation through a
virtual gluon mode at lowest order, the produced $Q\bar Q$ pair is in
a color-octet state.  The corrective factor to be used is the
color-octet corrective factor, $K_{q\bar q}= K({\rm color~octet})$ for
the quark-gluon-plasma.  The corrected cross-section is then
\begin{eqnarray}
\sigma_{q\bar q} (M_{Q\bar Q}) =  K_{q\bar q} { 8 \pi\alpha^2_s \over 27 M_{Q\bar
Q}^2}
\biggl (1+{\eta_m
\over 2}\biggr) \sqrt{1-\eta_m}\,.
\end{eqnarray}

In $Q\bar Q$ production by gluon fusion, $Q\bar Q$ 
pairs are produced with a relative color-octet and
color-singlet population given by \cite{Fad90}
\begin{eqnarray}
{ {\rm (color-octet)} \over {\rm (color-singlet)}}=
{(d^{\rm{abc}}/\sqrt{2})^2 \over (\delta^{\rm{ab}}/\sqrt{3})^2} ={5\over 2}
\end{eqnarray}

Taking the weights of the population into account, we can
write the corrective factor for the gluon fusion mode as $K_{gg}=
[5K({\rm octet}) +2K({\rm singlet})]/7$.
The cross-section for the gluon fusion production
of $Q\bar Q$ pairs, corrected for the effect of screening is
then
\begin{eqnarray}
\sigma_{gg} (M_{Q\bar Q}) = K_{gg} { \pi\alpha^2_s \over 3 M_{Q\bar
Q}^2}\biggl\{ (1 +\eta_m +{1\over 16} \eta_m^2) \ln
\biggl({1+\sqrt{1-\eta_m}\over 1-\sqrt{1-\eta_m}} \biggr) - \biggl(
{7\over 4} + {31\over16} \eta_m \biggr) \sqrt{1-\eta_m}
\biggr)\biggr\}\,.
\end{eqnarray}

In Figures 9 and 10 we present the cross sections for $c\bar c$ and $s
\bar s$ production corrected for screening effects in the quark-gluon
plasma, together with the uncorrected cross sections for comparison.
The cross sections for the case with the color-Coulomb interaction
without screening are also shown.  For $s\bar s$ and $c\bar c$
production by $q\bar q$ annihilation, the intermediate virtual gluon
selects color-octet states only. The color interaction is therefore
repulsive and suppresses the cross section.  With typical screening
lengths of 0.2 and 0.4 fm, the corrected cross sections are near the
no-screening color-Coulomb values for $c \bar c$, while for $s \bar s$
production they are close to the tree level lowest-order values in
accordance with the different values of $\eta$ for the two flavors.

The gluon fusion modes on the other hand produce $q\bar q$ pairs in a
combination of color-singlet and color-octet states, and the larger
magnitude of the color-singlet $K$-factor dominates the correction
even though the singlet weight factor is lower. Here again, the cross
sections for $c\bar c$ production approach the no-screening
color-Coulomb limit, whereas the cross sections for $s\bar s$
production remain close to the lowest-order values, due to the
differences in $\eta$. It may also be remarked that $c\bar c$
production cross sections through the gluon fusion mode for
$\lambda_{{}_D}=$ 0.4 fm exhibit a marked rise at small velocities
caused by the maximum in the $K$-factor occurring as the lowest bound
state emerges into the continuum at $\eta=$ 0.835.  On the other hand,
the $s\bar s$ production cross section is not appreciably affected by
the corrective factors in the presence of screened fields. Distortion
corrections in such a field are not expected to influence $s\bar s$
cross sections in the plasma noticeably.

\section{Conclusions and Discussions}

In reactions involving the production or the annihilation of a quark
and an antiquark, the quark and the antiquark are subject to
initial-state or final-state interactions due to their mutual
interactions.  The lowest-order cross sections for these processes can
be corrected by using an approximate corrective $K$-factor to take
into account the $q$-$\bar q$ interaction.  In a deconfined
quark-gluon plasma, the interaction between the quark and the
antiquark is screened.  As a consequence, the corresponding $K$-factor
will be modified.  We have studied the effect of screening on the
corrective $K$-factor.  Our knowledge of such an effect can be used to
search for the quark-gluon plasma.  If the corrective $K$-factor can
be inferred from experimental data, then we can use the relation
between the color screening and the $K$-factor to deduce the
conditions of screening and the possible presence of the deconfined
phase of strongly-interacting matter.

In the present investigation we have limited our attention to energies
near the $q$-$\bar q$ threshold, where the dominant contribution to
the corrective $K$-factor comes from the distortion of the wave
function by the $q$-$\bar q$ interaction which is modified by
screening.  The $K$-factors for the case with screening have been
determined by numerical integration of the non-relativistic
Schr\"odinger equation for $q$-$\bar q$ with a color-Yukawa potential.
The $K$-factor gives an enhancement for color-singlet $q\bar q$
states, due to the attractive nature of the color-singlet potential,
but a suppression for color-octet $q\bar q$ states due to the
repulsive nature of the color-octet interaction. The color-singlet
enhancement is larger in magnitude than the color-octet suppression
because of the difference in the two strengths specified by the color
factors of $4/3$ and $-1/6$ respectively.

The corrective $K$-factor for the Yukawa interaction, $-\alpha_{\rm
eff}e^{-r/\lambda_D}/r$, depends on two dimensionless parameters: the
usual `Coulomb' parameter $\xi=\alpha_{\rm eff}/v$, and the screening
length parameter $\eta=\lambda_{{}_D}/a_{{}_B}$.  We have evaluated
$K(\xi,\eta)$ for a large range of of $\xi$ and $\eta$, to study how
the $K$-factor depends on the $q$-$\bar q$ relative velocity and the
screening length.  The function $K(\xi,\eta)$ expressed in the
dimensionless parameters $\xi$ and $\eta$ can be used for other
reactants interacting through a Yukawa potential.

For attractive Yukawa potentials we observe prominent peaks of the
$K$-factor as a function of the screening length parameter $\eta$.
The peaks are located at $\eta=0.835$ and $\eta=3.23$, corresponding
to lowest two $s$-wave $q \bar q$ bound states emerging into the
continuum to become $q\bar q$ resonances as the screening length
decreases.  These $q\bar q$ resonances can be called ``screening
resonances'' as they arise from the screening of the $q$-$\bar q$
interaction in the quark-gluon plasma.  We have shown that their
behavior is compatible with resonances expected for potentials with
finite range. This is also in accord with results obtained for the
scattering of an electron from an atom \cite{Mor32,All31}, and is
similar to the manifestation of resonances through the $s$-wave
strength functions in low-energy neutron scattering from nuclei
\cite{Boh67}. The explicit location of the peak at $\lambda_{{}_D}=
0.835 a_{{}_B}$ coincides with the screening length at which the
lowest $q\bar q$ state becomes unbound\cite{Mat88}.

We have calculated the $K$-factors for two typical choices of the
Debye screening length ($0.2$ fm and $0.4$ fm), corresponding to
plasma temperatures of approximately $400$ and $200$ MeV respectively.
They are given separately for color-singlet and color-octet $q$-$\bar
q$ states and can be used to correct the lowest-order cross
sections. We have used these $K$-factors to investigate the effects of
screening on the cross sections for charm and strange pair production.
While the corrections to the cross section in the no-screening
color-Coulomb limit are of similar magnitude for $s\bar s$ and $c\bar
c$ pairs in the same velocity range, they are considerably different
for the two systems in the presence of screening with a color-Yukawa
interaction.  This arises because the Bohr radius is much smaller for 
the $c$-$\bar c$ system than for the $s$-$\bar s$ system.  A screening
length of 0.4 fm would correspond to $\eta\approx 0.8$ for $c$-$\bar
c$ but only 0.08 for $s$-$\bar s$.  The screening length paramter
$\eta=0.8$ happens also to be near the location where a $c$-$\bar c$
resonance occurs at $E\approx 0$.  Thus, $c$-$\bar c$ color singlet
production is much enhanced for $\lambda=0.4$ fm.  There is no such
enhancement for $s$-$\bar s$ production at this screening length.

Since the single virtual gluon in $q\bar q$ annihilation leads to
color octet pair production, the lowest order rates for $q\bar
q\rightarrow c\bar c (s\bar s)$ are larger than the corrected ones
because the octet correction is suppressive.  On the other hand,
$q\bar q$ production by gluon fusion can proceed through color-singlet
or octet states, the weight being given by $2/5$ for singlet to
octet. Despite the lower weight, the larger magnitude of the
color-singlet $K$-factor results in the attractive potential
dominating, so that the net effect of the correction is an enhancement
of gluon fusion production.

It is also noteworthy that the peak of the $K$-factor in the $c\bar c$
system corresponds to the unbinding of the lowest bound state and its
emergence as a $c\bar c$ resonance.  It is manifested as a rise in the
gluon fusion cross sections near threshold, for a screening length of
0.4 fm, which is near the screening length required for a peak in the
$K$-factor, as shown in Fig.\ 9.  Gluon fusion being the dominant mode
for heavy-quark production, the combined effect is to have a peak in
the $c\bar c$ production cross section just above the $c\bar c$
threshold, with a relatively narrow width in energy.  Thus, the
occurrence of a $c\bar c$ resonance will be signalled by a large
enhancement of $c \bar c$ production just above the threshold.  This
is in contrast to $s \bar s$ production for $\lambda_{{}_D}=0.2-0.4$
fm, for which the corresponding screening length parameters of
$\eta=0.04-0.08$ are very small and are far from $\eta=0.835$ for a
$q\bar q$ resonance.

From our results, the occurrence of $q\bar q$ screening resonances may
provide a way to search for the quark-gluon plasma.  The $q\bar q$
resonances give rise to prominent peaks of the $K$-factor as a
function of the screening length parameter, which is the ratio of the
screening length $\lambda_{{}_D}$ to the Bohr radius $a_{{}_B}$.  The
screening length is inversely proportional to the temperature
\cite{Gro81}.  Thus, $q\bar q$ resonances lead to prominent peaks of
the $K$-factor as a function of the plasma temperature.  We have seen
in Fig.\ 9 that large values of the $K$-factor near the threshold give
rise a narrow peak in the heavy-quark production cross section just
above the threshold.  The occurrence of a $q \bar q$ resonance will be
accompanied by a much enhanced $q\bar q$ production cross section just
above the threshold, and the enhancement will be a function of the
temperature.

The search for $q\bar q$ screening resonances in the quark-gluon
plasma can make use of the peaks of the $K$-factor at $\eta=0.835$ and
$\eta=$3.23.  The resonance at $\eta=3.23$ may not lead to realizable
enhancements because it corresponds to temperatures much below the
estimated quark-gluon plasma transition temperature (of approximately
$150-200$ MeV).  Using the perturbative QCD estimates, the screening
length parameter of $\eta=0.835$ corresponds to a $c$-$\bar c$
resonance at $T_{c\bar c} \approx 165$ MeV and a $b \bar b$ resonance
at $T_{b\bar b}\approx 393$ MeV.  These $T_{c\bar c}$ and $T_{b\bar
b}$ estimates from PQCD are approximate and uncertain, as lattice
gauge theory gives Debye screening lengths of about half of PQCD
estimates.  (See Fig.\ 7 of \cite{Uka89}.)  The Debye screening length
needs to be determined by experimental searches for these $c\bar c$
and $b\bar b$ resonances using the peaks in the $K$-factors.  The PQCD
estimates are useful only as a rough guide.  Thus, for $T\approx
T_{c\bar c}$ (approximately 165 MeV), and if the quark-gluon plasma is
produced, $\eta(c\bar c)=0.835$ and there will be a large enhancement
of $c\bar c$ production near the threshold, much greater than what one
expects from the lowest-order cross sections (Fig.\ 3).  At
temperatures far from the $c\bar c$ resonance temperature $T_{c\bar
c}$ there is no such large enhancement of $c\bar c$ production near
threshold.  At $T\approx T_{b\bar b}$ (approximately 393 MeV), if the
quark-gluon plasma is produced, $\eta(b\bar b)=0.835$ and there will
be a large enhancement of $b\bar b$ production near the threshold,
again much greater than expected from the lowest-order cross sections.
At temperatures far from the $b\bar b$ resonance temperature $T_{b\bar
b}$, there is no such large enhancement of $b\bar b$ production near
the thresholds.  Temperature dependence of this type arises from the
nature of screening between the interacting heavy quark and its
antiquark partner, which is an important property to identify the
deconfined quark-gluon plasma.  A search for the quark-gluon plasma
using heavy-quark resonances will require the measurement of the
production yield of heavy-quark pairs near the threshold, and a method
to estimate the temperature of the environment in which the production
takes place.  The enhancement will occur either for the production of
heavy-quark pairs by the collision of the constituents of the
thermalized quark-gluon plasma, or by the collision of the partons in
nucleon-nucleon collisions in a quark-gluon plasma environment.

It is interesting to note the recent report from the NA38 and the
Helios-3 Collaborations of an excess of dileptons at a dilepton
invariant mass of 1.5 to 2.5 GeV \cite{Lou95,Ram95,Mas95}.  The NA38
Collaboration suggests that this may be due to an excess production of
charm pairs in high-energy nucleus-nucleus collisions \cite{Ram95}.
The estimate from our work gives $c\bar c$ screening resonance
occurring at a temperature $T_{c\bar c}$ about 165 MeV, which is
within the experimental range of temperatures as encountered in the
NA38 and the Helios-3 experiments.  It will be of great interest to
have a direct measurement of charm pair production near the threshold
for the reactions studied by NA38 and Helios-3, to see whether there
are indeed excess charm pairs near the threshold, and whether these
excess charm pairs arise as a manifestation of a $c\bar c$ resonance
emerging just above the threshold in a quark-gluon plasma, or from
some other processes \cite{Won96b}.

After the work in the present paper was completed, a recent study of
the effect of a repulsive screened Coulomb interaction between two
pions in pion interferometry was brought to our attention\cite{Anc96}.
The modification of the Gamow factor due to screening was obtained in
\cite{Anc96} by using a quasi-classical approximation to barrier
penetration, whereas the results obtained here (Section III) for
repulsive screening potential follows from exact integration of the
Schr\"odinger equation.  For this special case of a repulsive screened
potential, our exact results complement the quasi-classical
approximate solutions of \cite{Anc96}.
  
\section{acknowledgements}
The authors would like to thank Dr. Ray Satchler, ORNL, for valuable
discussions.  Lali Chatterjee would like to thank Dr. Anand K. Bhatia,
NASA, Goddard Space Flight Center for helpful discussions, and Dr. M.
Strayer and Dr. F. Plasil for their kind hospitality at ORNL.  This
research was supported in part by the Division of Nuclear Physics,
U.S. Department of Energy under Contract No. DE-AC05-96OR22464, managed
by Lockheed Martin Energy Research Corp.

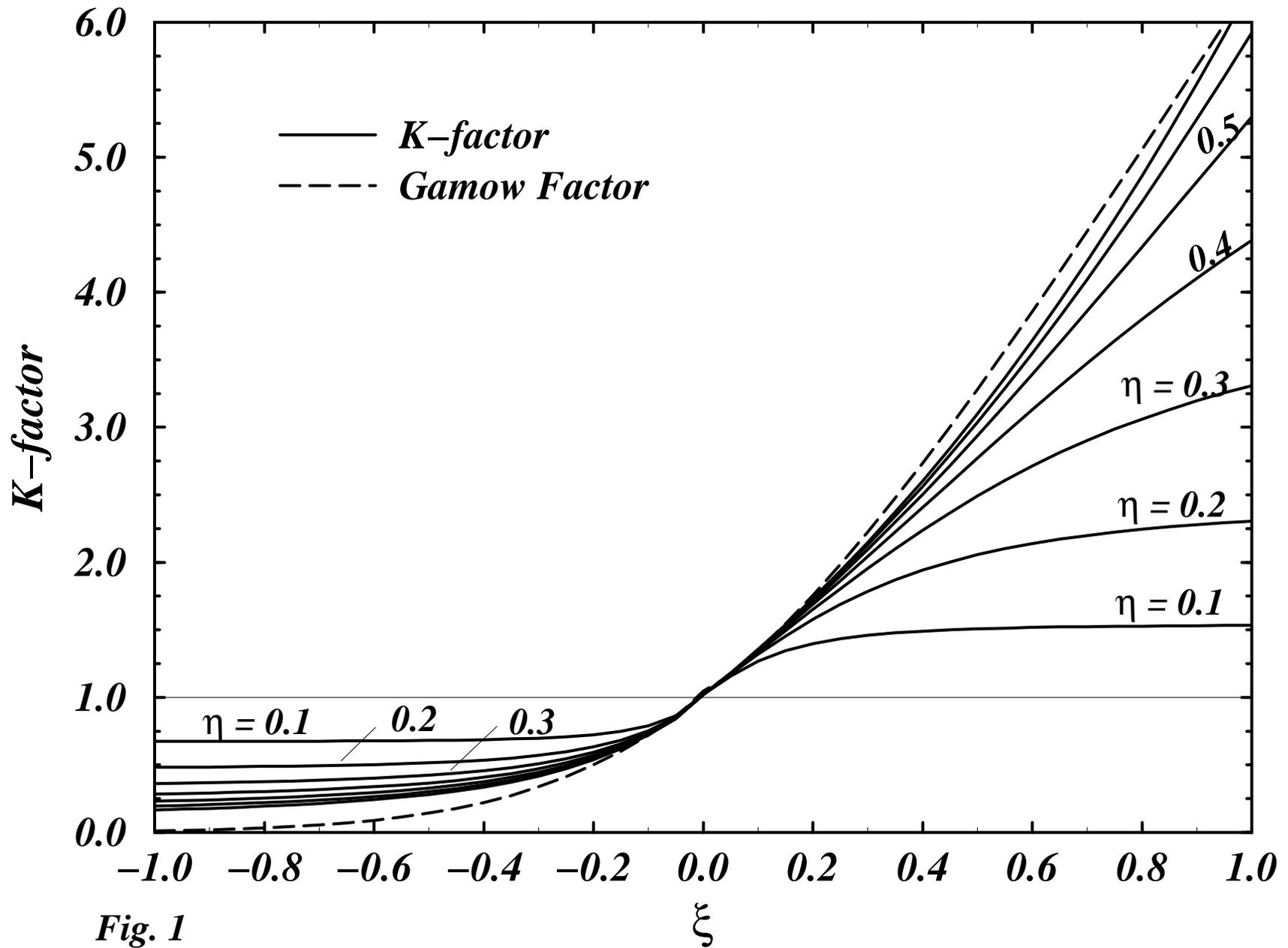
\begin{figure}[htbp]
 \caption{The corrective factor $K(\xi,\eta)$ plotted as a function of
$\xi$ for different values of the dimensionless screening parameter
$\eta$.  The parameter $\xi$ is positive for an attractive Yukawa
potential and is negative for a repulsive Yukawa potential.  The Gamow
factor for the case of Coulomb (or color-Coulomb) interaction is also
shown as the dashed curve. } \label{fig1}
\end{figure}

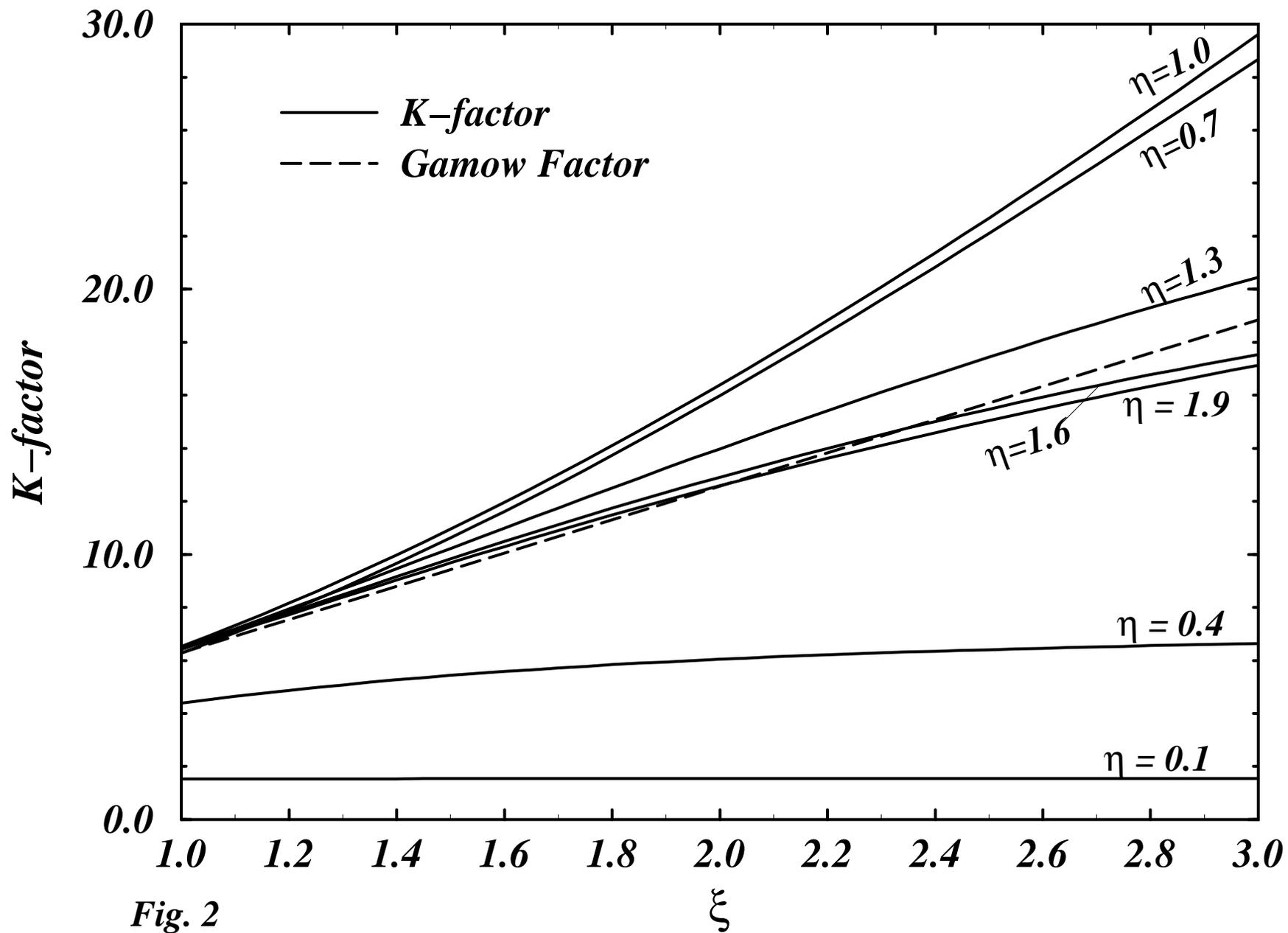
\begin{figure}[htbp]
\caption{The corrective factor $K(\xi,\eta)$ plotted as a function of
$\xi$ for $1\le \xi \le 3$ and various values of $\eta$.  The Gamow
factor for the case of Coulomb interaction is also shown as the dashed
curve. } \label{fig2}
\end{figure}

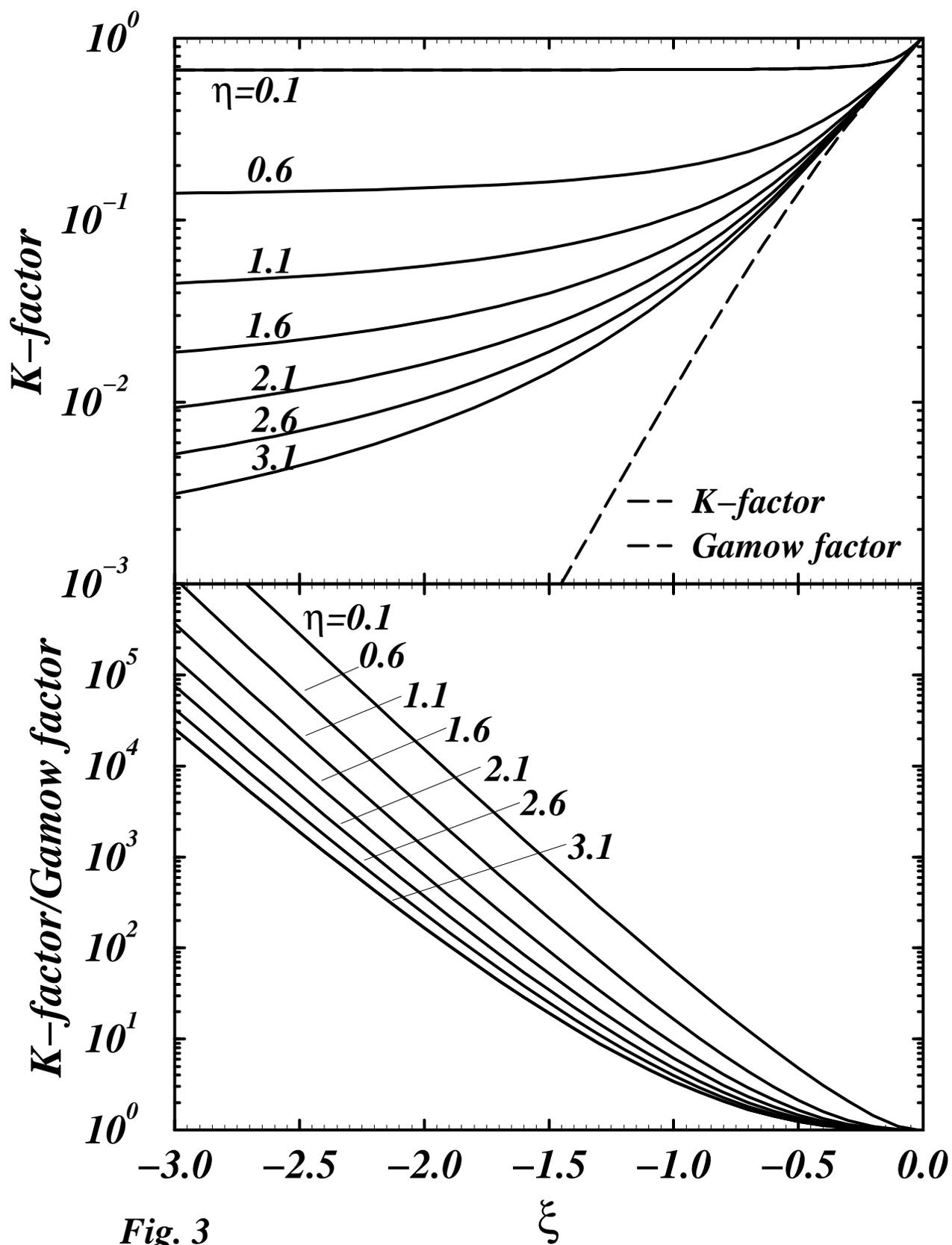
\begin{figure}[htbp]
\caption{(a) The corrective factor $K(\xi,\eta)$ plotted as a
function of $\eta$ for $0.4\le \eta \protect\le 2$ and different
values of $\xi$, (b) the ratio $K(\xi,\eta)$/(Gamow factor) as a
function $\protect\eta$ for different values of
$\protect\xi$.}\label{fig3}
\end{figure}

\noindent
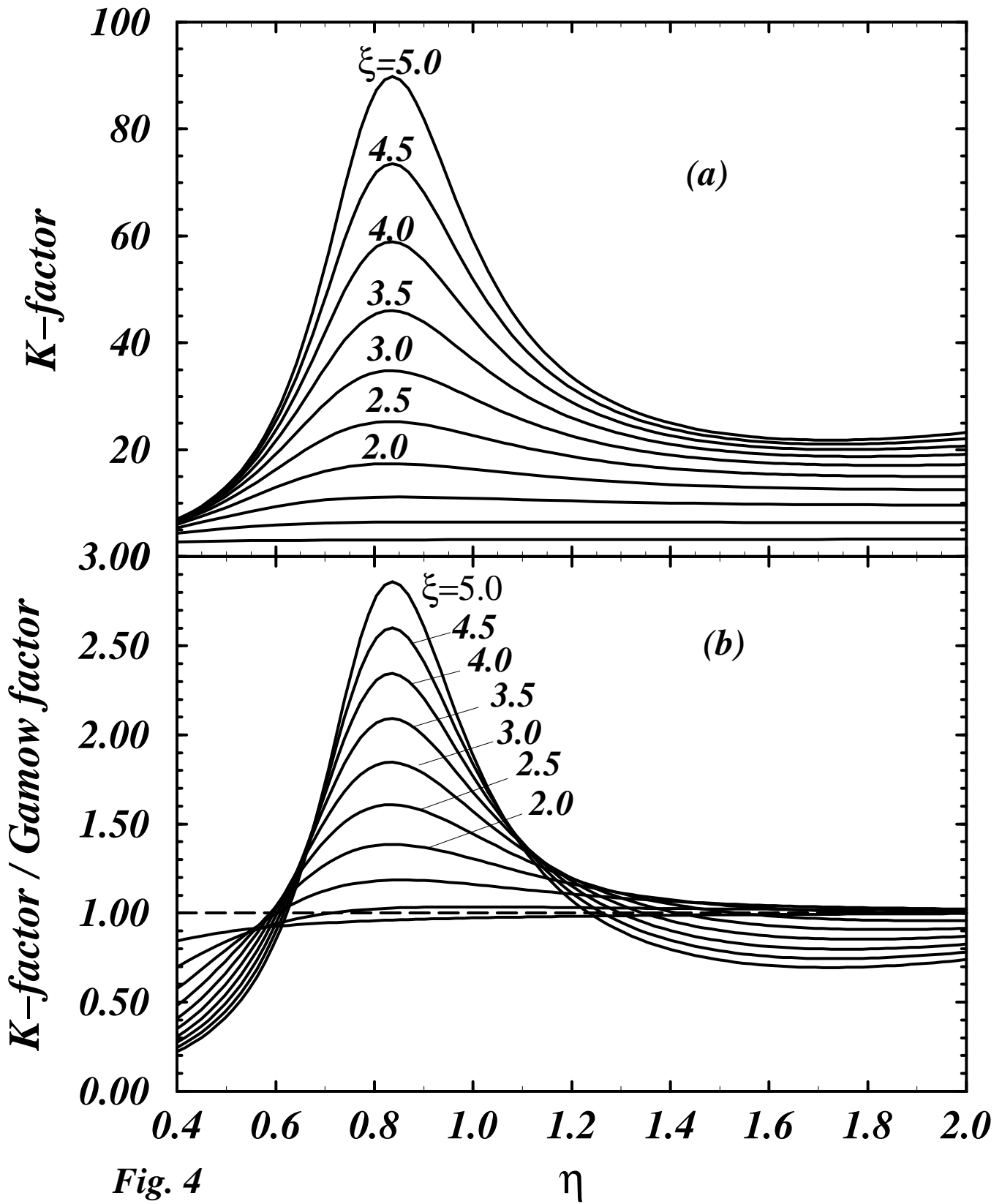
\begin{figure}
\caption{The phase shift $\protect\delta_0$ as a function of energy for
$\protect\eta$ near 0.835.}\label{fig4}
\end{figure}

\begin{figure}[htbp]
\caption{(a) The corrective factor $K(\xi,\eta)$ plotted as a
function of $\eta$ for $1.5\le \eta\le 5.0$ and different values of
$\xi$, (b) the ratio $K(\xi,\eta)$/(Gamow factor) as a function $\eta$
for different values of $\xi$.} \label{fig5}
\end{figure}

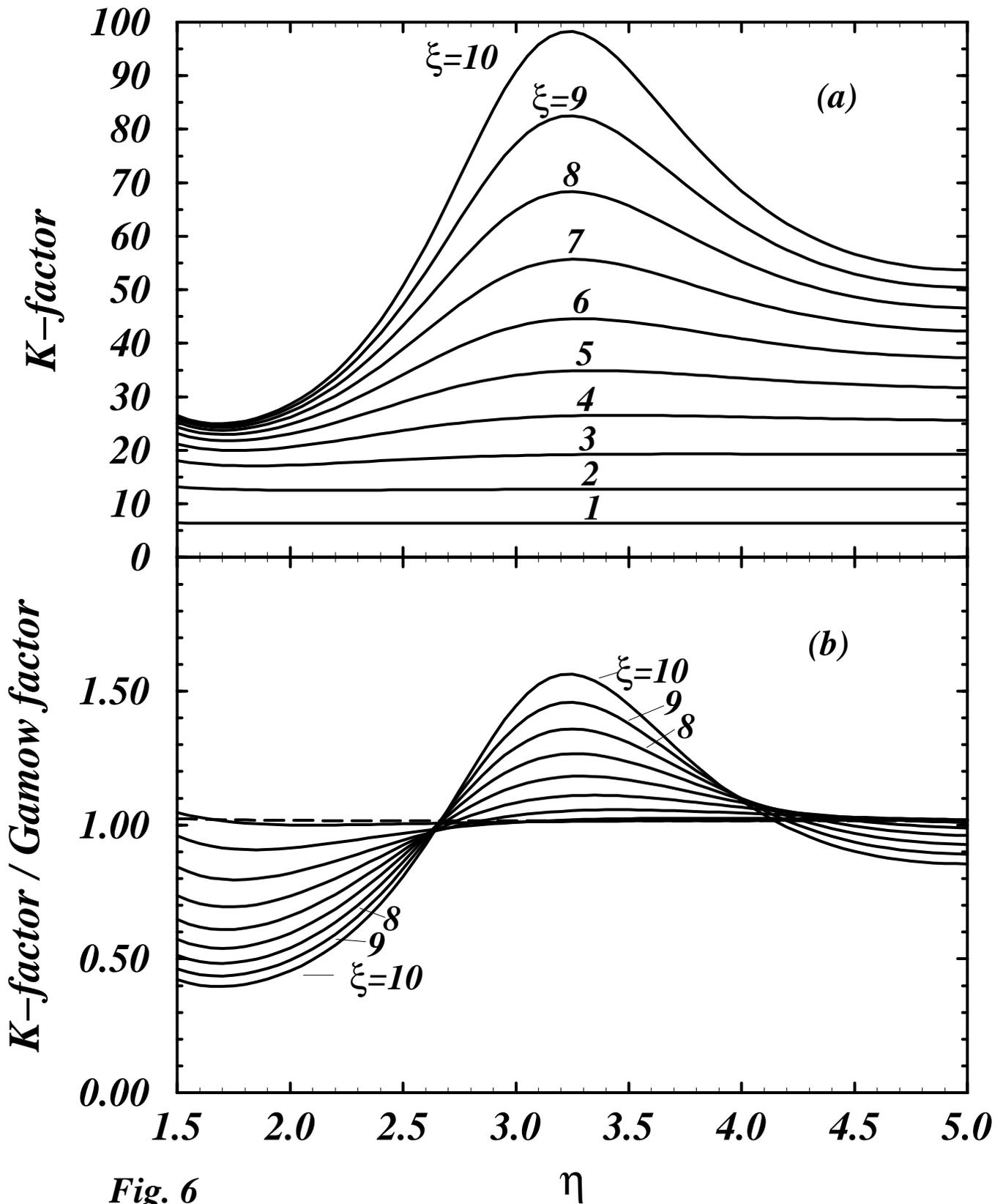
\begin{figure}[htbp]
\caption{(a) The corrective factor $K(\xi,\eta)$ plotted as a function
of $\xi$ for $-3\le \xi \le 0$ and different values of $\eta$.
Negative values of $\xi$ represent a repulsive Yukawa interaction. The
Gamow factor for the case of Coulomb interaction is shown as the
dashed curve.  (b) the ratio $K(\xi,\eta)$/(Gamow factor) as a
function $\xi$ for different values of $\eta$.} \label{fig6}
\end{figure}

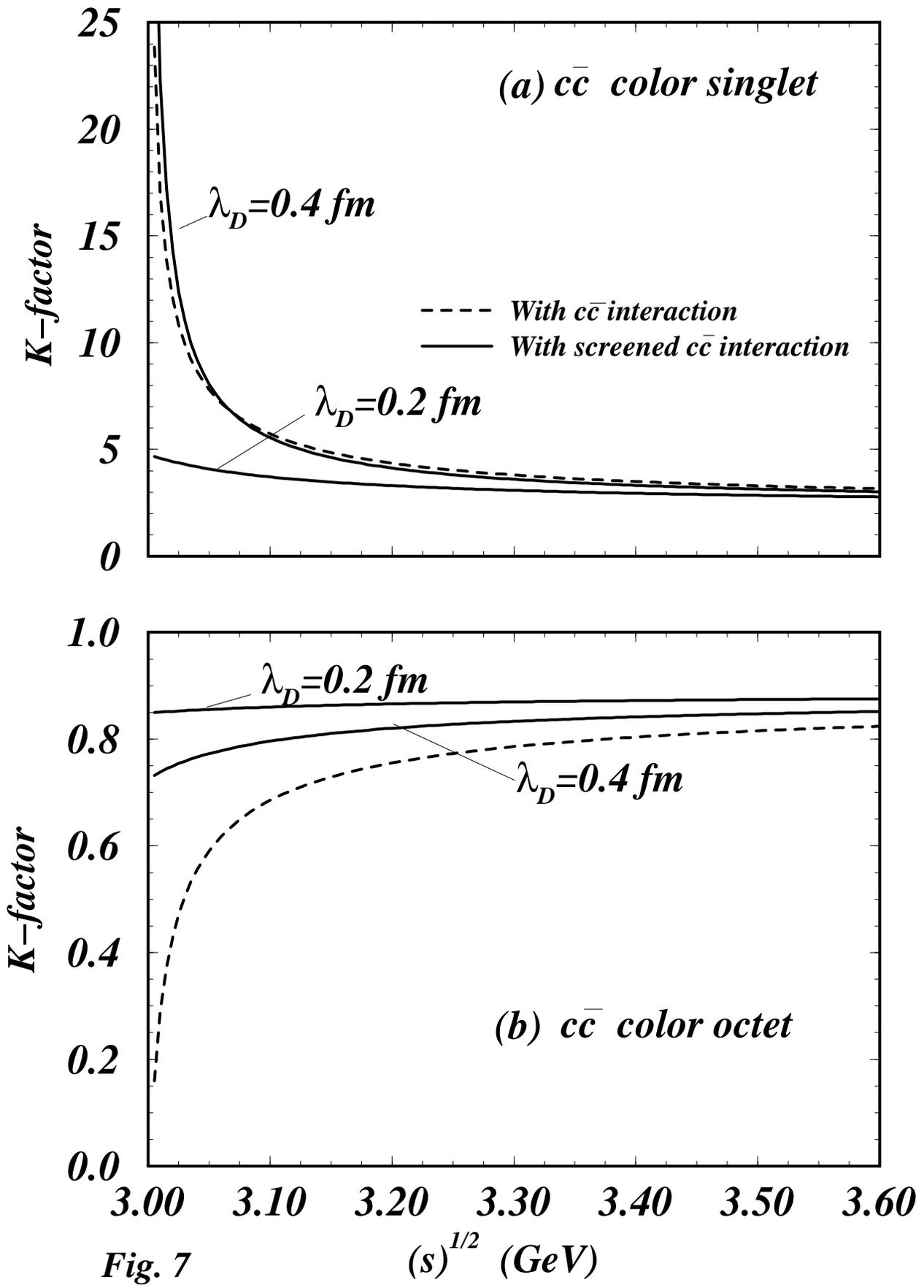
\begin{figure}[htbp]
\caption{The corrective factor $K(\xi,\eta)$ for a $c\protect\bar c$
pair with screening length $\lambda_{{}_D}$=0.2 and 0.4 fm, as a
function of the center-of-mass energy $\protect\sqrt{s}$. The Gamow
factor is shown as a dashed curve for comparison. Fig. 7(a) is for
color-singlet $c\bar c$ states, and Fig. 7(b) is for color-octet
$c\bar c$ states.}\label{fig7}
\end{figure}

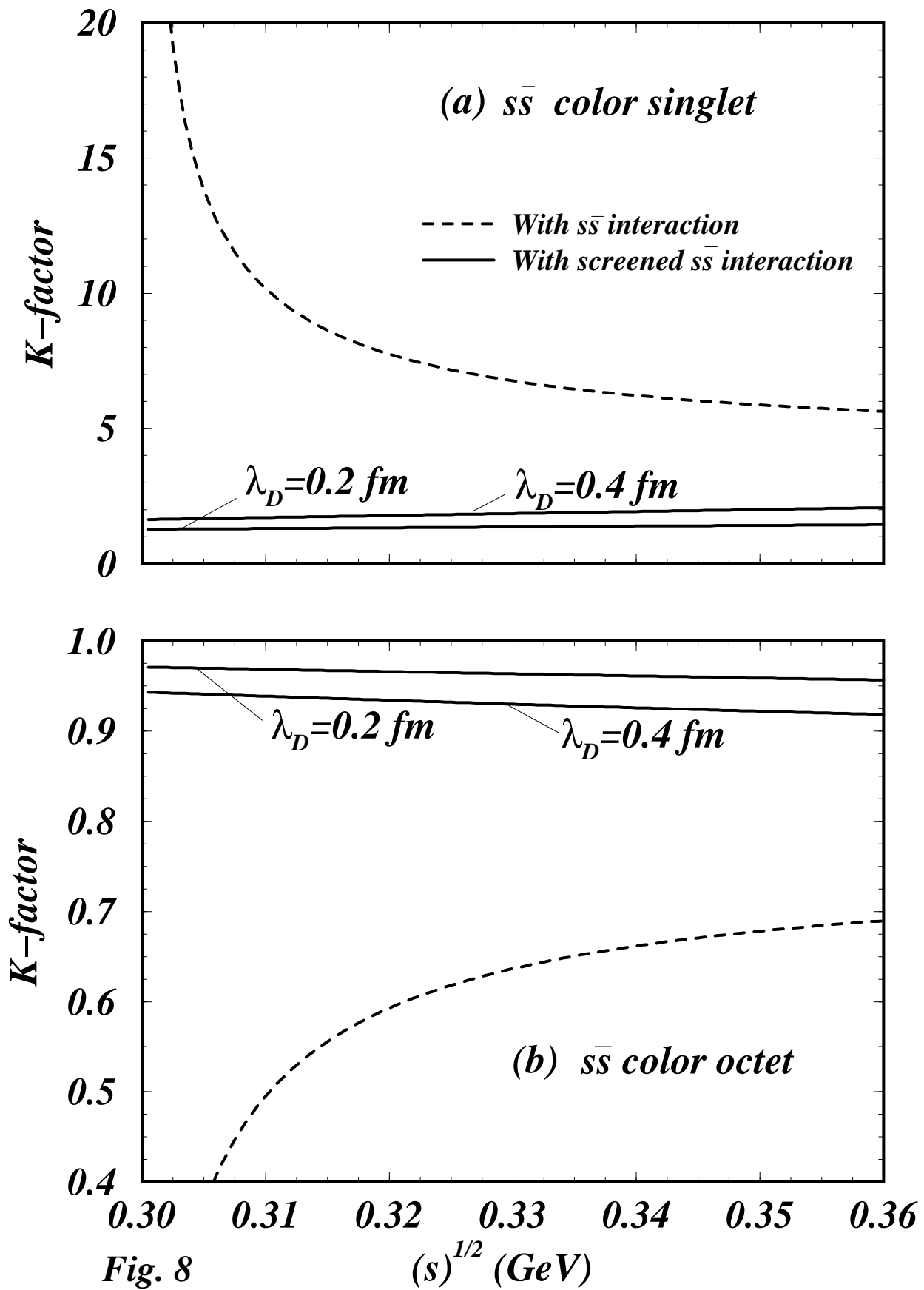
\begin{figure}[htbp]
\caption{The corrective factor $K(\xi,\eta)$ for an $s\bar s$ pair
with screening length $\lambda_{{}_D}$=0.2 and 0.4 fm, as a function
of the center-of-mass energy $\protect\sqrt{s}$. The Gamow factor is
shown as a dashed curve for comparison. Fig. 8(a) is for color-singlet
$s \bar s$ states, and Fig. 8(b) is for color-octet $s \bar s$
states.}\label{fig8}
\end{figure}

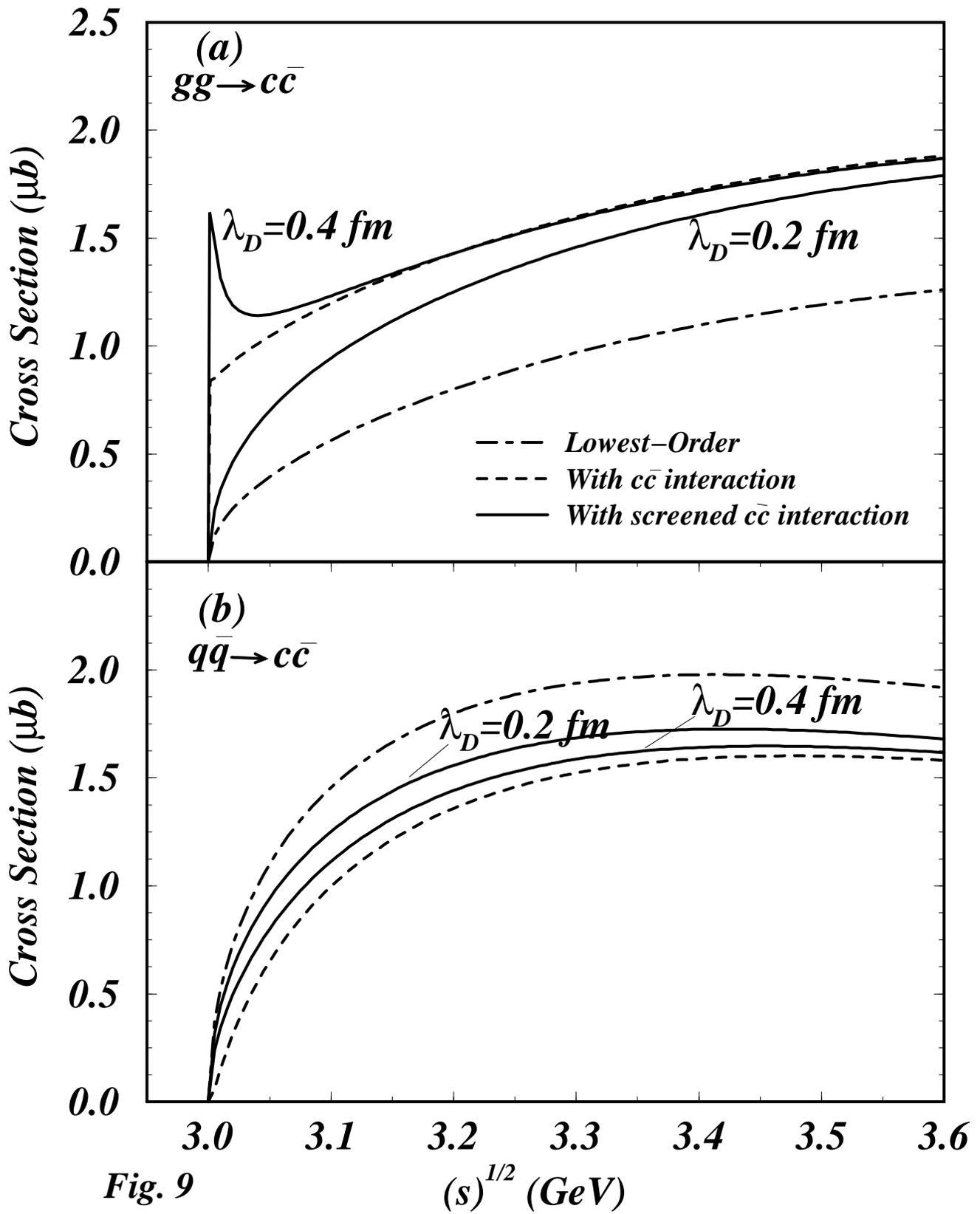
\begin{figure}[htbp]
\caption{Cross sections for $c \bar c$ production (a) by gluon fusion,
and (b) by $q\bar q$ annihilation, with screening lengths of
$\lambda_{{}_D}=0.2$ and $0.4$ fm. The lowest-order cross section and
the cross section in the case of color-Coulomb $c$-$\bar c$
interaction are also shown.}\label{fig9}
\end{figure}

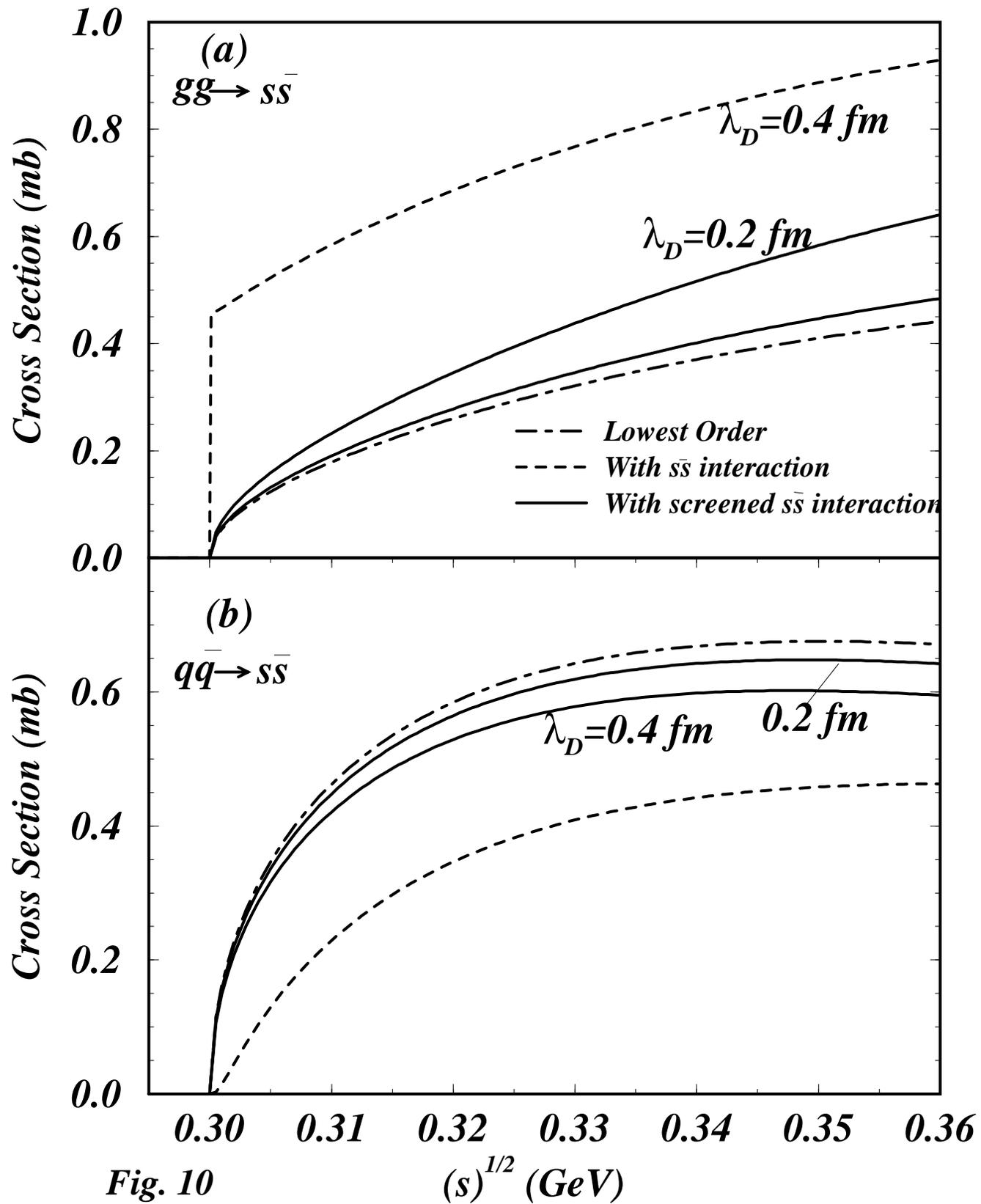
\begin{figure}[htbp]
\caption{Cross sections for $s \bar s$ production (a) by gluon fusion,
and (b) by $q\bar q$ annihilation, with screening lengths of
$\lambda_{{}_D}=0.2$ and $0.4$ fm. The lowest-order cross section and the cross
section in the case of color-Coulomb $s$-$\bar s$ interactions are
also shown.}\label{fig10}
\end{figure}

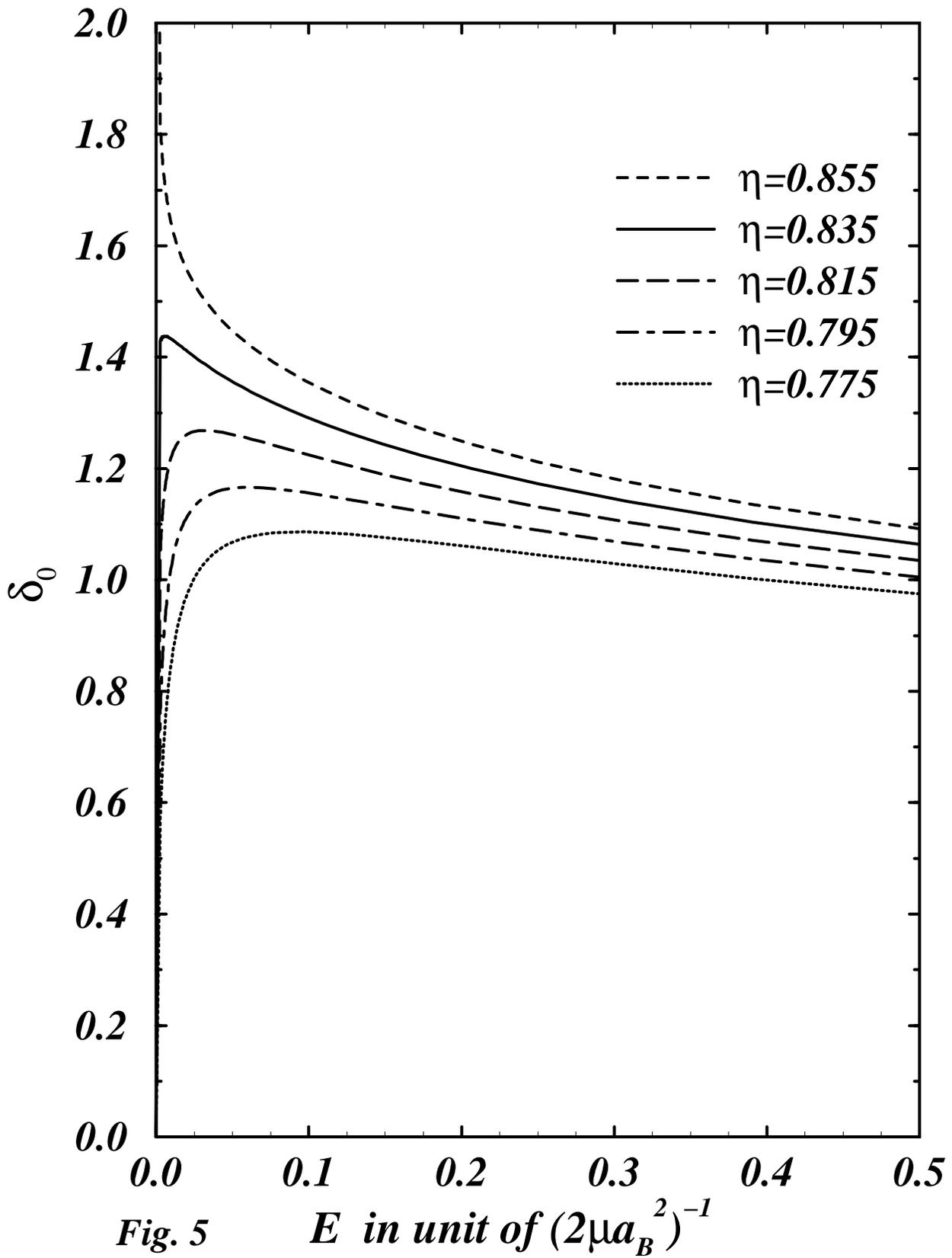
\end{document}